\documentclass[lettersize,journal]{IEEEtran}
\usepackage{amsmath,amsfonts}
\usepackage{algorithmic}
\usepackage{algorithm}
\usepackage{array}
\usepackage{textcomp}
\usepackage{stfloats}
\usepackage{url}
\usepackage{verbatim}
\usepackage{graphicx}
\usepackage{cite}
\usepackage{multirow}

\usepackage{tikz}
\usepackage{pgfplots}
\usetikzlibrary{arrows,shapes,circuits.logic.US,shapes.gates.logic.IEC,calc,decorations.markings,patterns,positioning,fit}

\usepackage{pgf-pie}

\usepackage{subcaption}

\usepackage{threeparttable}

\usepackage{fancyhdr}
\makeatletter
\def\ps@IEEEtitlepagestyle{
    \def\@oddfoot{\strut\hfill\parbox{\textwidth}{This work has been submitted to the IEEE for possible publication. Copyright may be transferred without notice, after which this version may no longer be accessible.}\hfill}%
    \def\@evenfoot{}%
    \def\@oddhead{}%
    \def\@evenhead{}%
}
\makeatother

\usepackage{url}

\usepackage{lipsum}

\tikzset{every node/.style={font=\sffamily}} 

\def\BibTeX{{\rm B\kern-.05em{\sc i\kern-.025em b}\kern-.08em
    T\kern-.1667em\lower.7ex\hbox{E}\kern-.125emX}}
\begin{document}
\bstctlcite{IEEEexample:BSTcontrol}
\title{L-Sort: On-chip Spike Sorting with Efficient Median-of-Median Detection and Localization-based Clustering}
\author{Yuntao Han, \IEEEmembership{Graduate Student Member, IEEE}, Yihan Pan, \IEEEmembership{Member, IEEE}, Xiongfei Jiang, \IEEEmembership{Graduate Student Member, IEEE}, Cristian Sestito, \IEEEmembership{Member, IEEE}, Shady Agwa, \IEEEmembership{Member, IEEE}, Themis Prodromakis, \IEEEmembership{Senior Member, IEEE}, and Shiwei Wang, \IEEEmembership{Senior Member, IEEE}
\thanks{This work was supported by the EPSRC Programme Grant FORTE under Grant EP/R024642/1, and the RAEng Chair in Emerging Technologies under Grant CiET1819/2/93, and the Royal Society under grant IEC/NSFC/223067. (Corresponding author: Yuntao Han)}
\thanks{Yuntao Han, Yihan Pan, Xiongfei Jiang, Cristian Sestito, Shady Agwa, Themis Prodromakis and Shiwei Wang are with the Centre for Electronics Frontiers, Institute for Integrated Micro and Nano Systems, School of Engineering, University of Edinburgh, Edinburgh, EH8 9YL, UK (e-mail: \{Yuntao.Han, yihan.pan, xiongfei.jiang, csestito, shady.agwa, t.prodromakis, shiwei.wang\}@ed.ac.uk).}
}

\maketitle

\begin{abstract}
Spike sorting is a critical process for decoding large-scale neural activity from extracellular recordings. The advancement of neural probes facilitates the recording of a high number of neurons with an increase in channel counts, arising a higher data volume and challenging the current on-chip spike sorters. This paper introduces L-Sort, a novel on-chip spike sorting solution featuring median-of-median spike detection and localization-based clustering. By combining the median-of-median approximation and the proposed incremental median calculation scheme, our detection module achieves a reduction in memory consumption. Moreover, the localization-based clustering utilizes geometric features instead of morphological features, thus eliminating the memory-consuming buffer for containing the spike waveform during feature extraction. Evaluation using Neuropixels datasets demonstrates that L-Sort achieves competitive sorting accuracy with reduced hardware resource consumption. Implementations on FPGA and ASIC (180 nm technology) demonstrate significant improvements in area and power efficiency compared to state-of-the-art designs while maintaining comparable accuracy. If normalized to 22 nm technology, our design can achieve roughly $\times 10$ area and power efficiency with similar accuracy, compared with the state-of-the-art design evaluated with the same dataset. Therefore, L-Sort is a promising solution for real-time, high-channel-count neural processing in implantable devices.
\end{abstract}


\section{Introduction}
\label{sec:introduction}
\IEEEPARstart{A}{cquiring} large-scale single-neuron activities is paramount in neuroscientific studies for understanding the brain~\cite{ss}. Implantable neural probes have been a prevailing device for attaining extracellular recordings. These recordings contain the superimposition of spikes generated by nearby neurons, along with background activities in the brain. These spikes represent the activations of their respective firing neurons. To discern the individual neuron activities from the recording, spike sorting is utilized to detect and attribute the spikes to their putative neurons.

Presently, modern neural probes (e.g., Neuropixels series~\cite{np1,np2,npultra}) are equipped with hundreds of recording sites (electrodes) with pitches down to a few micrometers. To keep up with the increase in data volume, on-chip spike sorting~\cite{suevey_quiroga} is a promising solution precluding the transmission of huge amounts of recording and meeting the stringent combination of processing capability and power efficiency for protecting the brain tissue, as well as facilitating closed-loop applications in a low-latency fashion.

\begin{figure}[t]
  \centering
  \includegraphics[width=.9\linewidth]{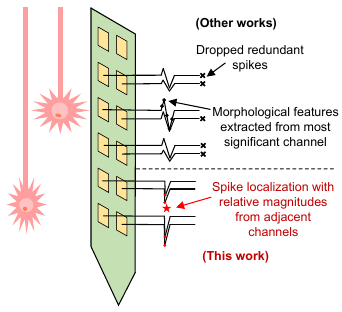}
  \caption{Morphological features v.s. spike localization (spatial features) in spike sorting.}
  \label{fig:intro}
\end{figure}

Compared with the offline spike sorting algorithms performed after-the-fact on power-hungry workstations with CPUs and GPUs, on-chip spike sorters are designed to conduct processing on-the-fly in a power- and area-constrained scenario. Therefore, unlike the offline designs utilizing computation- and memory-intensive algorithms iteratively for maximizing the sorting accuracy, the on-chip spike sorting commonly uses a three-step pipeline~\cite{ah_survey} to decipher the spikes from the recording, namely, spike detection, feature extraction, and clustering.

Spike detection is the first step in the pipeline~\cite{detection_ic}, performed on the recordings from the probe which is preprocessed by digital filtering. While there are several designs on software exploring accurate detection with neural networks~\cite{yass} or template matching~\cite{ks4}, the existing on-chip spike sorter commonly utilized simpler methods, e.g., thresholding, to find the outliers (spikes). The thresholds are either set as a fixed absolute value~\cite{li2018esscirc,han2021tbcas,fereshteh2022tvlsi} or calculated dynamically using previous samples during the sorting process~\cite{tuan2019tvlsi,zeinolabedin2022tbcas,jssc_2023}. Dynamic thresholds are promising for achieving higher accuracy, as they can adapt to time-variant conditions and are better suited for multi-channel recordings, avoiding the labor-intensive process of fine-tuning fixed thresholds for each channel. However, since spike detection is an always-on module on hardware, whether spikes are present or not, it is necessary that the dynamical threshold calculation is implemented in a hardware-efficient manner for achieving a low-power design. Mean and median are statistical measures commonly used to determine thresholds, and are often implemented using accumulators and comparators, respectively. Medians are generally preferred over means because they are less affected by outliers (spikes)~\cite{detection_ic}. However, medians are less efficient in hardware because the comparison between each sample with each sample is required, i.e., $O(n^2)$ computational complexity.

\begingroup
\renewcommand{\arraystretch}{1.2}
\begin{table}[t]
\centering
\caption{Works of Multi-channel Spike Sorting Using different Features}
\label{table:intro}
\begin{threeparttable}
\setlength\tabcolsep{1.5pt}
\begin{tabular}{lllll}
\hline
\textbf{Work}                        & \textbf{Platform} & \textbf{Detection} & \textbf{Feature}     & \textbf{Clustering} \\ \hline
NIPS'2019~\cite{localization_nips19} & GPU               & -                  & 2D-geometry          & AVI                 \\
NIPS'2021~\cite{localization_nips21} & GPU               & -                  & 3D-geometry$^1$      & point-cloud         \\
NIPS'2024~\cite{bypassing_nips2024}    & GPU               & fixed-TH           & 2D-geometry          & -                   \\ \hline
TBCAS'2019~\cite{posort}             & FPGA/ASIC         & mean-TH            & -                    & O-Sort              \\
Access'2020~\cite{zyon}              & FPGA              & amplitude-TH       & FSDE                 & K-Means             \\
TBME'2020~\cite{geo_osort_fpga}      & FPGA              & mean-TH            & -                    & O-Sort$^2$          \\
TBCAS'2023~\cite{detection_ic}       & FPGA/ASIC         & median-TH          & -                    & -                   \\
JSSC'2023~\cite{jssc_2023}           & ASIC$^3$              & mean-TH            & peak-FSDE            & O-Sort$^2$          \\
TBCAS'2024~\cite{tbcas_2024}         & FPGA/ASIC         & mean-TH            & FSDE                 & O-Sort              \\ \hline
\textbf{This work}                   & \textbf{FPGA/ASIC}     & \textbf{median-TH} & \textbf{2D-geometry} & \textbf{O-Sort}     \\ \hline
\end{tabular}
\begin{tablenotes}
    \item[1] with an additional dimension calculated with triangulation.
    \item[2] explored locality with geometry information.
    \item[3] evaluated on silicon.
\end{tablenotes}
\end{threeparttable}

\end{table}
\endgroup

The following two steps, feature extraction and clustering, attribute the detected spikes to different neurons. Because the waveform of each detected spike can spread in many timesteps and channels, the reduction of dimensionality for each spike is necessary for reducing computation during clustering. Conventionally, the spikes recorded by low-channel-count probes with distantly placed electrodes are typically picked up by only a single channel. By contrast, modern probes are equipped with tightly placed electrodes, between which the pitch is merely few micrometers. Indicatively, the spikes are sensed by multiple adjacent electrodes, thus providing spatial features like location information of their putative neurons, as shown in Fig.~\ref{fig:intro}.  Spike localization~\cite{localization_ori} is a recently emerging topic that infers the geometric position of the spike source from multi-channel recordings with electrode geometry, as shown in TABLE~\ref{table:intro}. The calculated geometric information could be utilized for clustering in the next step. As the spikes could be captured by multiple electrodes, the localization could be performed by assessing the relative voltage amplitudes of these channels. A widely used technique for spike localization is the center of mass (CoM) method~\cite{localization_ori}, which has shown good accuracy in distinguishing spike sources~\cite{localization_cell}. This method computes the source position by taking the weighted average of the positions of a selected set of channels. The weights are the amplitudes of these channels, typically including all surrounding channels centered around the one with the highest amplitude. Compared with the calculation of conventional features, e.g., first-and-second-derivative-extrema (FSDE)~\cite{geo_osort_fpga,jssc_2023}, involving the whole recording in the time window, these positions can be calculated with very few timesteps, eliminating the need for retaining all the recorded signals in the time window. Our recent work, L-Sort, explored efficient hardware implementation utilizing modified spike localization techniques~\cite{l_sort}, which achieved lower memory utilization and access compared with the designs using morphological methods because of the preclusion of memory-consuming on-chip storage of whole spike waveforms.

In this paper, we further improve the hardware efficiency of L-Sort, especially aiming to reduce memory footprint by using a novel scheme for approximated median-of-median-based peak detection and a further simplified spike localization process. We compared the FPGA testing results with the figures reported in our previous paper, and it shows that further improvement on hardware efficiency has been achieved by these technologies. Besides, we implemented the architecture on ASIC using a standard 180 nm CMOS technology and compared the results with existing state-of-the-art on-chip spike sorters.

The rest of this paper is organized as follows: Section~\ref{sec:method} discusses the overall architecture of L-Sort and details the optimization techniques for achieving more hardware-efficient implementation, with respect to detection and sorting accuracy; Section~\ref{sec:experiments} demonstrates the sorting results of the proposed hardware design and reports the hardware utilization on both FPGA and ASIC, with comparisons with relative state-of-the-art designs; Section~\ref{sec:conclusion} summarizes this work. 

\section{Methodology}
\label{sec:method}

\begin{figure*}[t]
  \centering
  \includegraphics[width=.8\linewidth]{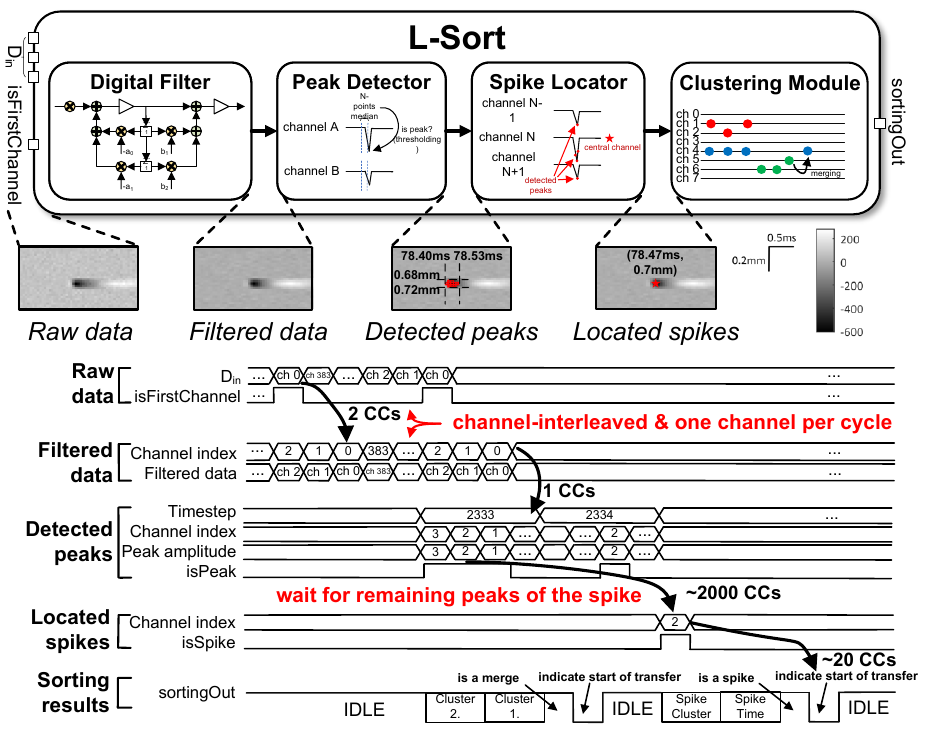}
  \caption{Overall hardware architecture of L-Sort.}
  \label{fig:overview}
\end{figure*}

\subsection{Overview}

The hardware design of L-Sort comprises a digital filter, peak detector, spike locator, and clustering module, as shown in Fig.~\ref{fig:overview}.

The input to our designed system is channel-interleaved and one sample from one channel is sent to the system in each cycle. For reducing the number of IOs, a single-bit port, \emph{isFirstChannel}, for indicating the transfer of the first channel instead of a multi-bit port for conveying channel indices along with the ports for carrying digitalized voltage values are implemented. The digital filter, as the first module of L-Sort, removes the local field potentials and high-frequency noises from the raw data. The filtered data is then processed by the peak detector to find the peaks in different channels based on median thresholding. These peaks are grouped into spikes and the locations of these spikes are calculated as geometric features. Finally, the clustering module utilizes these positions to classify these spikes into different neurons with the O-Sort algorithm. Considering the intrinsic sparsity of spikes in electrophysiology recordings, only a one-bit port, \emph{sortingOut}, is utilized for outputting the sorting results. \emph{sortingOut} remains high when there is no result to be transferred. Once a spike is detected and attributed to a cluster (putative neuron), or a merging between two clusters is triggered, \emph{sortingOut} will become zero for one cycle, indicating the start of the transfer. In the next cycle, \emph{sortingOut} will indicate whether the result is a sorted spike or a cluster merging with a ‘1’ or ‘0’, which is followed by the timestamp and cluster index of the spike or the indices of two merged clusters, respectively.

\subsection{Digital Filter}

The digital filter is a first-order infinite-impulse-response (IIR) filter with band-pass frequency from $300$ to $6000$ Hz. The implementation of this filter follows the Direct Form \uppercase\expandafter{\romannumeral2}~\cite{wp330}, whose coefficients are quantized to $12$-bit signed and fixed representation with $10$ bits for fractions.

\subsection{Median-of-Median-based Spike Detection}
The peak detector finds samples with relatively high absolute values from the filtered data through channel-wise median thresholding, whose hardware implementation requires both the median calculation and the comparison between the calculated threshold and the sample. The threshold $TH[c,t]$ could be calculated as follows:
\begin{equation}
\begin{aligned}
    TH[c,t]&=N_{th}\times median(x[c,t-N:t])
\end{aligned}
\end{equation}
where $c$ and $t$ are channel indices and timestamps, $N$ is the number of points used for calculating median. The value of $N$ has significant impacts on the detection accuracy and hardware utilization. A higher value provides a more accurate estimation of the signal statistics and therefore achieves a higher accuracy. However, the median calculation requires comparing every point with the rest of the points, resulting in an $O(n^2)$ algorithm complexity. Consequently, a higher $N$ requires more hardware resources in implementation. In this paper, $N$ is set to $25$, considering the optimal trade-off between accuracy and hardware complexity.

\begin{figure}[t]
    \centering
    \begin{subfigure}[b]{\linewidth}
        \centering
        \scalebox{.8}{
            \begin{tikzpicture}[]

\def\boxsizeX{1}
\def\boxsizeY{0.125}
\def\offsetX{0}
\def\offsetY{0}

\def\boxsizeMX{1.3}
\def\boxsizeMY{.5}

\def\boxsizeNX{1}
\def\boxsizeNY{.5}

\foreach \y in {0,...,24} {
    \draw[draw=black] (\offsetX,\offsetY+\boxsizeY*\y) rectangle (\offsetX+\boxsizeX,\offsetY+\boxsizeY*\y+\boxsizeY);
}

\draw[->] (\offsetX-1.4, \offsetY+0.5*\boxsizeY) -- (\offsetX, \offsetY+0.5*\boxsizeY)
    node[pos=0, above] {newest sample} node[pos=0, below] {$X[N-1]$};

\draw[<-] (\offsetX-1.4, \offsetY+24*\boxsizeY+0.5*\boxsizeY) -- (\offsetX, \offsetY+24*\boxsizeY+0.5*\boxsizeY)
    node[pos=0, above] {oldest sample} node[pos=0, below] {$X[0]$};

\draw[<-,blue,thick] (\offsetX-.4, \offsetY+24*\boxsizeY+0.5*\boxsizeY-0.5) -- (\offsetX-.4, \offsetY+0.5*\boxsizeY+0.5)
    node[midway, left] {\color{blue}\textbf{\begin{tabular}{cc}stored in \\ order of \\ timestep\end{tabular}}};

\draw[->] (\offsetX+\boxsizeX, \offsetY+12*\boxsizeY+0.5*\boxsizeY) -- (\offsetX+\boxsizeX+.5, \offsetY+12*\boxsizeY+0.5*\boxsizeY);

\draw[draw=black] (\offsetX+\boxsizeX+.5,\offsetY+12*\boxsizeY+0.5*\boxsizeY-.5*\boxsizeMY) rectangle (\offsetX+\boxsizeX+.5+\boxsizeMX,\offsetY+12*\boxsizeY+0.5*\boxsizeY+.5*\boxsizeMY);

\node at (\offsetX+\boxsizeX+.5+0.5*\boxsizeMX, \offsetY+12*\boxsizeY+0.5*\boxsizeY) {Median};

\draw[->] (\offsetX+\boxsizeX+.5+\boxsizeMX, \offsetY+12*\boxsizeY+0.5*\boxsizeY) -- (\offsetX+\boxsizeX+.5+\boxsizeMX+.5, \offsetY+12*\boxsizeY+0.5*\boxsizeY);

\draw[draw=black] (\offsetX+\boxsizeX+.5+\boxsizeMX+.5,\offsetY+12*\boxsizeY+0.5*\boxsizeY-.5*\boxsizeMY) rectangle (\offsetX+\boxsizeX+.5+\boxsizeMX+.5+\boxsizeNX,\offsetY+12*\boxsizeY+0.5*\boxsizeY+.5*\boxsizeMY);

\node at (\offsetX+\boxsizeX+.5+\boxsizeMX+.5+0.5*\boxsizeNX, \offsetY+12*\boxsizeY+0.5*\boxsizeY) {$\times N_{th}$};

\draw[->] (\offsetX+\boxsizeX+.5+\boxsizeMX+.5+\boxsizeNX, \offsetY+12*\boxsizeY+0.5*\boxsizeY) -- (\offsetX+\boxsizeX+.5+\boxsizeMX+.5+\boxsizeNX+.5, \offsetY+12*\boxsizeY+0.5*\boxsizeY);

\end{tikzpicture}
        }
        \caption{Conventional Median Calculator.}
        \label{fig:median_a}
    \end{subfigure}
    \begin{subfigure}[b]{\linewidth}
    \vspace{10pt}
        \centering
        \scalebox{.8}{
            \begin{tikzpicture}[]

\def\boxsizeX{1}
\def\boxsizeY{0.125}
\def\offsetX{0}
\def\offsetY{0}

\def\boxsizeMX{1.3}
\def\boxsizeMY{.5}

\def\boxsizeNX{1}
\def\boxsizeNY{.5}

\foreach \my in {0,...,4} {
    \foreach \y in {0,...,4} {
        \draw[draw=black] (\offsetX,\offsetY+\boxsizeY*\y+6*\boxsizeY*\my) rectangle (\offsetX+\boxsizeX,\offsetY+\boxsizeY*\y+\boxsizeY+6*\boxsizeY*\my);
    }
}

\draw[->] (\offsetX-1.4, \offsetY+0.5*\boxsizeY) -- (\offsetX, \offsetY+0.5*\boxsizeY)
    node[pos=0, above] {newest sample} node[pos=0, below] {$X[N-1]$};

\draw[<-] (\offsetX-1.4, \offsetY+28*\boxsizeY+0.5*\boxsizeY) -- (\offsetX, \offsetY+28*\boxsizeY+0.5*\boxsizeY)
    node[pos=0, above] {oldest sample} node[pos=0, below] {$X[0]$};

\draw[<-,blue,thick] (\offsetX-.4, \offsetY+28*\boxsizeY+0.5*\boxsizeY-0.5) -- (\offsetX-.4, \offsetY+0.5*\boxsizeY+0.5)
    node[midway, left] {\color{blue}\textbf{\begin{tabular}{cc}stored in \\ order of \\ timestep\end{tabular}}};

\foreach \my in {0,...,4} {
    \draw[->] (\offsetX+\boxsizeX, \offsetY+2*\boxsizeY+0.5*\boxsizeY+6*\boxsizeY*\my) -- (\offsetX+\boxsizeX+.5, \offsetY+2*\boxsizeY+0.5*\boxsizeY+6*\boxsizeY*\my);
    
    \draw[draw=black] (\offsetX+\boxsizeX+.5,\offsetY+2*\boxsizeY+0.5*\boxsizeY-.5*\boxsizeMY+6*\boxsizeY*\my) rectangle (\offsetX+\boxsizeX+.5+\boxsizeMX,\offsetY+2*\boxsizeY+0.5*\boxsizeY+.5*\boxsizeMY+6*\boxsizeY*\my);
    
    \node at (\offsetX+\boxsizeX+.5+0.5*\boxsizeMX, \offsetY+2*\boxsizeY+0.5*\boxsizeY+6*\boxsizeY*\my) {Median};
}

\draw[->] (\offsetX+\boxsizeX+.5+\boxsizeMX, \offsetY+2*\boxsizeY+0.5*\boxsizeY) -- (\offsetX+\boxsizeX+.5+\boxsizeMX+.7, \offsetY+11*\boxsizeY+0.5*\boxsizeY);
\draw[->] (\offsetX+\boxsizeX+.5+\boxsizeMX, \offsetY+8*\boxsizeY+0.5*\boxsizeY) -- (\offsetX+\boxsizeX+.5+\boxsizeMX+.5, \offsetY+11*\boxsizeY+0.5*\boxsizeY);
\draw[->] (\offsetX+\boxsizeX+.5+\boxsizeMX, \offsetY+14*\boxsizeY+0.5*\boxsizeY) -- (\offsetX+\boxsizeX+.5+\boxsizeMX+.5, \offsetY+14*\boxsizeY+0.5*\boxsizeY);
\draw[->] (\offsetX+\boxsizeX+.5+\boxsizeMX, \offsetY+20*\boxsizeY+0.5*\boxsizeY) -- (\offsetX+\boxsizeX+.5+\boxsizeMX+.5, \offsetY+17*\boxsizeY+0.5*\boxsizeY);
\draw[->] (\offsetX+\boxsizeX+.5+\boxsizeMX, \offsetY+26*\boxsizeY+0.5*\boxsizeY) -- (\offsetX+\boxsizeX+.5+\boxsizeMX+.7, \offsetY+17*\boxsizeY+0.5*\boxsizeY);

\draw[draw=black] (\offsetX+\boxsizeX+.5+.5+\boxsizeMX,\offsetY+14*\boxsizeY+0.5*\boxsizeY-.5*\boxsizeMY) rectangle (\offsetX+\boxsizeX+.5+\boxsizeMX+.5+\boxsizeMX,\offsetY+14*\boxsizeY+0.5*\boxsizeY+.5*\boxsizeMY);

\node at (\offsetX+\boxsizeX+.5+0.5*\boxsizeMX+.5+\boxsizeMX, \offsetY+14*\boxsizeY+0.5*\boxsizeY) {Median};

\draw[->] (\offsetX+\boxsizeX+.5+\boxsizeMX+.5+\boxsizeMX, \offsetY+14*\boxsizeY+0.5*\boxsizeY) -- (\offsetX+\boxsizeX+.5+\boxsizeMX+.5+.5+\boxsizeMX, \offsetY+14*\boxsizeY+0.5*\boxsizeY);

\draw[draw=black] (\offsetX+\boxsizeX+.5+\boxsizeMX+.5+.5+\boxsizeMX,\offsetY+14*\boxsizeY+0.5*\boxsizeY-.5*\boxsizeMY) rectangle (\offsetX+\boxsizeX+.5+\boxsizeMX+.5+\boxsizeNX+.5+\boxsizeMX,\offsetY+14*\boxsizeY+0.5*\boxsizeY+.5*\boxsizeMY);

\node at (\offsetX+\boxsizeX+.5+\boxsizeMX+.5+0.5*\boxsizeNX+.5+\boxsizeMX, \offsetY+14*\boxsizeY+0.5*\boxsizeY) {$\times N_{th}$};

\draw[->] (\offsetX+\boxsizeX+.5+\boxsizeMX+.5+\boxsizeNX+.5+\boxsizeMX, \offsetY+14*\boxsizeY+0.5*\boxsizeY) -- (\offsetX+\boxsizeX+.5+\boxsizeMX+.5+\boxsizeNX+.5+.5+\boxsizeMX, \offsetY+14*\boxsizeY+0.5*\boxsizeY);

\end{tikzpicture}
        }
        \caption{Conventional Median-of-median Calculator.}
        \label{fig:median_b}
    \end{subfigure}
    \begin{subfigure}[b]{\linewidth}
    \vspace{10pt}
        \centering
        \scalebox{.8}{
            \begin{tikzpicture}[]

\def\boxsizeX{1}
\def\boxsizeY{0.125}
\def\offsetX{0}
\def\offsetY{0}

\def\boxsizeMX{1.3}
\def\boxsizeMY{.5}

\def\boxsizeNX{1}
\def\boxsizeNY{.5}

\foreach \y in {0,...,23} {
    \draw[draw=black] (\offsetX-.2,\offsetY+\boxsizeY*\y) rectangle (\offsetX,\offsetY+\boxsizeY*\y+\boxsizeY);
    \draw[draw=black] (\offsetX,\offsetY+\boxsizeY*\y) rectangle (\offsetX+\boxsizeX,\offsetY+\boxsizeY*\y+\boxsizeY);
}

\draw[->] (\offsetX-1.6, \offsetY+0.5*\boxsizeY) -- (\offsetX-.2, \offsetY+0.5*\boxsizeY)
    node[pos=0, above] {smallest sample};

\draw[->] (\offsetX-1.6, \offsetY+23*\boxsizeY+0.5*\boxsizeY) -- (\offsetX-.2, \offsetY+23*\boxsizeY+0.5*\boxsizeY)
    node[pos=0, below] {biggest sample};

\draw[<-,blue,thick] (\offsetX-.4, \offsetY+23*\boxsizeY+0.5*\boxsizeY-0.5) -- (\offsetX-.4, \offsetY+0.5*\boxsizeY+0.5)
    node[midway, left] {\color{blue}\textbf{\begin{tabular}{cc}stored in \\ order of \\ magnitude\end{tabular}}};

\draw[<-,red,thick] (\offsetX-.1, \offsetY) -- (\offsetX-.1, \offsetY-0.5)
    node[pos=1,left] {\color{red}\textbf{\begin{tabular}{c}extra memory usage \\ for marking timestep\end{tabular}}} node[pos=1,right] {\color{red}$log_2(N-1)$ bits per sample};

\draw[->] (\offsetX+\boxsizeX, \offsetY+12*\boxsizeY+0.5*\boxsizeY) -- (\offsetX+\boxsizeX+.5, \offsetY+12*\boxsizeY+0.5*\boxsizeY);

\draw[draw=black] (\offsetX+\boxsizeX+.5,\offsetY+12*\boxsizeY+0.5*\boxsizeY-.5*\boxsizeMY) rectangle (\offsetX+\boxsizeX+.5+\boxsizeMX,\offsetY+12*\boxsizeY+0.5*\boxsizeY+.5*\boxsizeMY);

\node at (\offsetX+\boxsizeX+.5+0.5*\boxsizeMX, \offsetY+12*\boxsizeY+0.5*\boxsizeY) {Median};

\draw[->] (\offsetX+\boxsizeX+.5+.5*\boxsizeMX,\offsetY+12*\boxsizeY+0.5*\boxsizeY-.5*\boxsizeMY-.5) -- (\offsetX+\boxsizeX+.5+.5*\boxsizeMX,\offsetY+12*\boxsizeY+0.5*\boxsizeY-.5*\boxsizeMY) node[pos=0,below] {\begin{tabular}{c}newest\\sample\end{tabular}};

\draw[->] (\offsetX+\boxsizeX+.5+.5*\boxsizeMX,\offsetY+12*\boxsizeY+0.5*\boxsizeY+.5*\boxsizeMY) -- (\offsetX+\boxsizeX+.5+.5*\boxsizeMX,\offsetY+12*\boxsizeY+0.5*\boxsizeY+.5*\boxsizeMY+.5) -- (\offsetX+\boxsizeX,\offsetY+12*\boxsizeY+0.5*\boxsizeY+.5*\boxsizeMY+.5) node[pos=.3,above] {update};

\draw[->] (\offsetX+\boxsizeX+.5+\boxsizeMX, \offsetY+12*\boxsizeY+0.5*\boxsizeY) -- (\offsetX+\boxsizeX+.5+\boxsizeMX+.5, \offsetY+12*\boxsizeY+0.5*\boxsizeY);

\draw[draw=black] (\offsetX+\boxsizeX+.5+\boxsizeMX+.5,\offsetY+12*\boxsizeY+0.5*\boxsizeY-.5*\boxsizeMY) rectangle (\offsetX+\boxsizeX+.5+\boxsizeMX+.5+\boxsizeNX,\offsetY+12*\boxsizeY+0.5*\boxsizeY+.5*\boxsizeMY);

\node at (\offsetX+\boxsizeX+.5+\boxsizeMX+.5+0.5*\boxsizeNX, \offsetY+12*\boxsizeY+0.5*\boxsizeY) {$\times N_{th}$};

\draw[->] (\offsetX+\boxsizeX+.5+\boxsizeMX+.5+\boxsizeNX, \offsetY+12*\boxsizeY+0.5*\boxsizeY) -- (\offsetX+\boxsizeX+.5+\boxsizeMX+.5+\boxsizeNX+.5, \offsetY+12*\boxsizeY+0.5*\boxsizeY);

\end{tikzpicture}
        }
        \caption{Incremental Median Calculator.}
        \label{fig:median_c}
    \end{subfigure}
    \begin{subfigure}[b]{\linewidth}
    \vspace{10pt}
        \centering
        \scalebox{.8}{
            \begin{tikzpicture}[]

\def\boxsizeX{1}
\def\boxsizeY{0.125}
\def\offsetX{0}
\def\offsetY{0}

\def\boxsizeMX{1.3}
\def\boxsizeMY{.5}

\def\boxsizeNX{1}
\def\boxsizeNY{.5}

\foreach \y in {0,...,3} {
    \draw[draw=black] (\offsetX-.2,\offsetY+\boxsizeY*\y) rectangle (\offsetX,\offsetY+\boxsizeY*\y+\boxsizeY);
    \draw[draw=black] (\offsetX,\offsetY+\boxsizeY*\y) rectangle (\offsetX+\boxsizeX,\offsetY+\boxsizeY*\y+\boxsizeY);
}

\draw[<-,blue,thick] (\offsetX-.4, \offsetY+3*\boxsizeY+0.5*\boxsizeY) -- (\offsetX-.4, \offsetY+0.5*\boxsizeY)
    node[midway, left] {\color{blue}\textbf{magnitude}};

\draw[->] (\offsetX+\boxsizeX, \offsetY+2*\boxsizeY+0.5*\boxsizeY) -- (\offsetX+\boxsizeX+.5, \offsetY+2*\boxsizeY+0.5*\boxsizeY);

\draw[draw=black] (\offsetX+\boxsizeX+.5,\offsetY+2*\boxsizeY+0.5*\boxsizeY-.5*\boxsizeMY) rectangle (\offsetX+\boxsizeX+.5+\boxsizeMX,\offsetY+2*\boxsizeY+0.5*\boxsizeY+.5*\boxsizeMY);

\node at (\offsetX+\boxsizeX+.5+0.5*\boxsizeMX, \offsetY+2*\boxsizeY+0.5*\boxsizeY) {Median};

\draw[->] (\offsetX+\boxsizeX+.5+.5*\boxsizeMX,\offsetY+2*\boxsizeY+0.5*\boxsizeY-.5*\boxsizeMY-.5) -- (\offsetX+\boxsizeX+.5+.5*\boxsizeMX,\offsetY+2*\boxsizeY+0.5*\boxsizeY-.5*\boxsizeMY) node[pos=0,below] {\begin{tabular}{c}newest\\sample\end{tabular}};

\draw[->] (\offsetX+\boxsizeX+.5+.5*\boxsizeMX,\offsetY+2*\boxsizeY+0.5*\boxsizeY+.5*\boxsizeMY) -- (\offsetX+\boxsizeX+.5+.5*\boxsizeMX,\offsetY+2*\boxsizeY+0.5*\boxsizeY+.5*\boxsizeMY+.3) -- (\offsetX+\boxsizeX-.5,\offsetY+2*\boxsizeY+0.5*\boxsizeY+.5*\boxsizeMY+.3) node[pos=.5,above] {update} -- (\offsetX+\boxsizeX-.5,\offsetY+2*\boxsizeY+0.5*\boxsizeY+.5*\boxsizeMY);

\draw[->] (\offsetX+\boxsizeX+.5+\boxsizeMX, \offsetY+2*\boxsizeY+0.5*\boxsizeY) -- (\offsetX+\boxsizeX+.5+\boxsizeMX+.5, \offsetY+2*\boxsizeY+0.5*\boxsizeY) -- (\offsetX+\boxsizeX+.5+\boxsizeMX+.5, \offsetY+2*\boxsizeY+0.5*\boxsizeY+1) -- (\offsetX+\boxsizeX+.5+.5*\boxsizeMX, \offsetY+2*\boxsizeY+0.5*\boxsizeY+1) -- (\offsetX+\boxsizeX+.5+.5*\boxsizeMX, \offsetY+2*\boxsizeY-0.5*\boxsizeMY+2);

\def\offsetX{0}
\def\offsetY{2}

\foreach \y in {0,...,3} {
    \draw[draw=black] (\offsetX-.2,\offsetY+\boxsizeY*\y) rectangle (\offsetX,\offsetY+\boxsizeY*\y+\boxsizeY);
    \draw[draw=black] (\offsetX,\offsetY+\boxsizeY*\y) rectangle (\offsetX+\boxsizeX,\offsetY+\boxsizeY*\y+\boxsizeY);
}

\draw[<-,blue,thick] (\offsetX-.4, \offsetY+3*\boxsizeY+0.5*\boxsizeY) -- (\offsetX-.4, \offsetY+0.5*\boxsizeY)
    node[midway, left] {\color{blue}\textbf{magnitude}};

\draw[->] (\offsetX+\boxsizeX, \offsetY+2*\boxsizeY+0.5*\boxsizeY) -- (\offsetX+\boxsizeX+.5, \offsetY+2*\boxsizeY+0.5*\boxsizeY);

\draw[draw=black] (\offsetX+\boxsizeX+.5,\offsetY+2*\boxsizeY+0.5*\boxsizeY-.5*\boxsizeMY) rectangle (\offsetX+\boxsizeX+.5+\boxsizeMX,\offsetY+2*\boxsizeY+0.5*\boxsizeY+.5*\boxsizeMY);

\node at (\offsetX+\boxsizeX+.5+0.5*\boxsizeMX, \offsetY+2*\boxsizeY+0.5*\boxsizeY) {Median};

\draw[->] (\offsetX+\boxsizeX+.5+.5*\boxsizeMX,\offsetY+2*\boxsizeY+0.5*\boxsizeY+.5*\boxsizeMY) -- (\offsetX+\boxsizeX+.5+.5*\boxsizeMX,\offsetY+2*\boxsizeY+0.5*\boxsizeY+.5*\boxsizeMY+.3) -- (\offsetX+\boxsizeX-.5,\offsetY+2*\boxsizeY+0.5*\boxsizeY+.5*\boxsizeMY+.3) node[pos=.5,above] {update every 5 timesteps} -- (\offsetX+\boxsizeX-.5,\offsetY+2*\boxsizeY+0.5*\boxsizeY+.5*\boxsizeMY);

\draw[->] (\offsetX+\boxsizeX+.5+\boxsizeMX, \offsetY+2*\boxsizeY+0.5*\boxsizeY) -- (\offsetX+\boxsizeX+.5+\boxsizeMX+.5, \offsetY+2*\boxsizeY+0.5*\boxsizeY);

\draw[draw=black] (\offsetX+\boxsizeX+.5+\boxsizeMX+.5,\offsetY+2*\boxsizeY+0.5*\boxsizeY-.5*\boxsizeMY) rectangle (\offsetX+\boxsizeX+.5+\boxsizeMX+.5+\boxsizeNX,\offsetY+2*\boxsizeY+0.5*\boxsizeY+.5*\boxsizeMY);

\node at (\offsetX+\boxsizeX+.5+\boxsizeMX+.5+0.5*\boxsizeNX, \offsetY+2*\boxsizeY+0.5*\boxsizeY) {$\times N_{th}$};

\draw[->] (\offsetX+\boxsizeX+.5+\boxsizeMX+.5+\boxsizeNX, \offsetY+2*\boxsizeY+0.5*\boxsizeY) -- (\offsetX+\boxsizeX+.5+\boxsizeMX+.5+\boxsizeNX+.5, \offsetY+2*\boxsizeY+0.5*\boxsizeY);

\end{tikzpicture}
        }
        \caption{Incremental Median-of-median Calculator.}
        \label{fig:median_d}
    \end{subfigure}
    \caption{Hardware architecture of Median calculators for 25 samples.}
\end{figure}
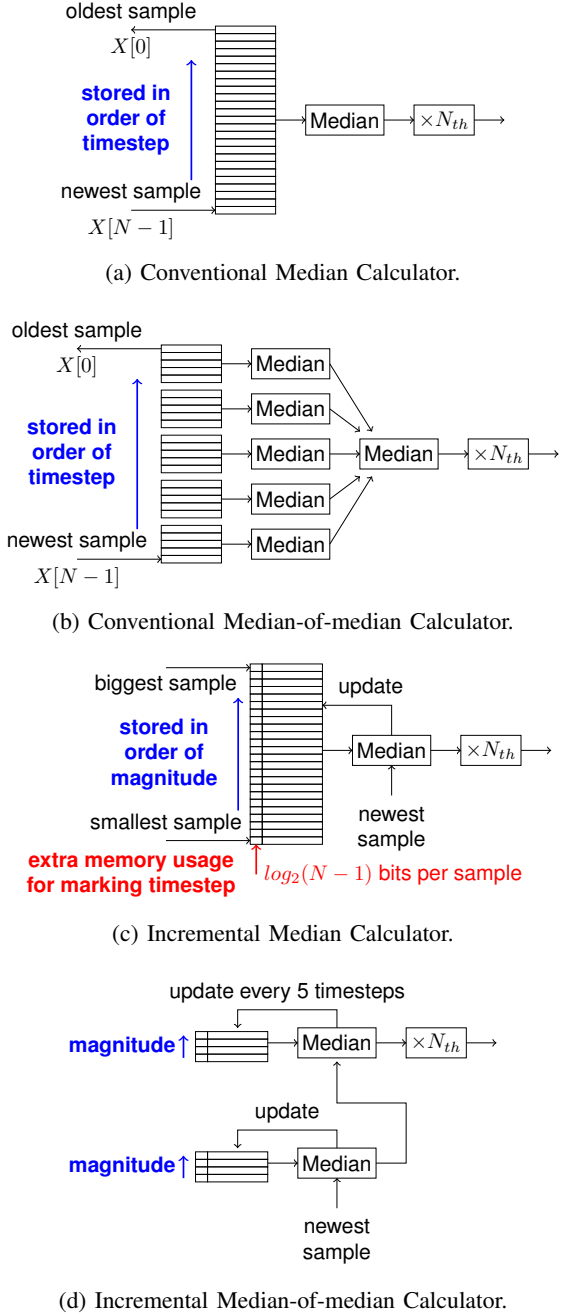

The hardware consumption and corresponding detection accuracy of different median calculators are shown in TABLE~\ref{table:exp_median} The conventional hardware implementation of a median calculator is shown in Fig.~\ref{fig:median_a}. A median buffer is implemented to store previous samples used to calculate the median for the current sample. These stored samples are ordered according to their timesteps. Each time a new sample is acquired, the oldest sample is discarded. However, because the sorting of all samples is performed in one clock cycle to find the median, this structure consumes more than $7500$ LUTs when implemented on FPGAs~\cite{detection_ic}. The median recursion (as shown in Fig.~\ref{fig:median_b}) has been proposed to reduce the logical hardware consumption, which estimated the real median with median-of-median to reduce the number of points involved in each median calculation. The rolled (shared median finder) and unrolled implementations of median-of-median calculators consume $468$ and $2312$ LUTs~\cite{detection_ic}, respectively. However, the memory consumption is unchanged since all the previous samples are still required during calculation.

\begin{figure}[t]
  \centering
  \includegraphics[width=.8\linewidth]{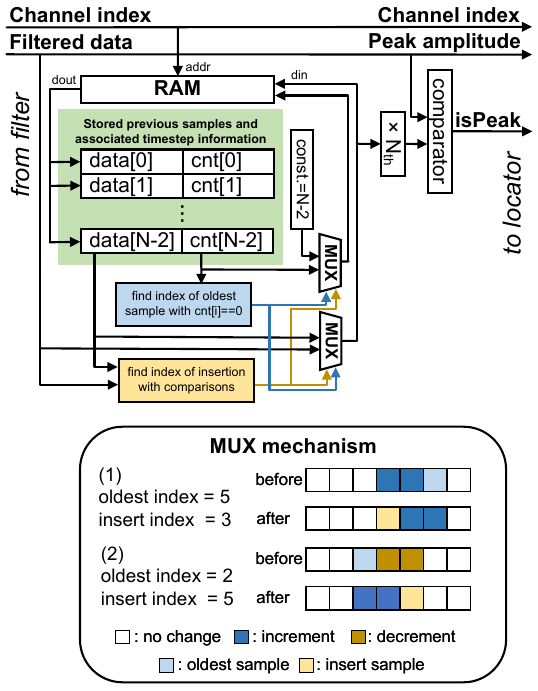}
  \caption{Hardware architecture for implementing incremental median calculation.}
  \label{fig:incrementaql_median}
\end{figure}

\begin{table}[t]
\renewcommand{\arraystretch}{1.3}
\setlength\tabcolsep{1.5pt}
\centering
\caption{Spike Detection Accuracy and Hardware Utilization of Different Median Calculators}
\label{table:exp_median}
\begin{tabular}{|c|c|cc|c|cc|}
\hline
\multirow{2}{*}{\textbf{Work}}               & \multirow{2}{*}{\textbf{Methodology}}                                       & \multicolumn{2}{c|}{\textbf{Acc.(\%)}}                               & \multirow{2}{*}{\textbf{Cycle}} & \multicolumn{2}{c|}{\textbf{HW Util.}}                    \\ \cline{3-4} \cline{6-7} 
                                             &                                                                             & \multicolumn{1}{c|}{set 1}                  & set 2                  &                                 & \multicolumn{1}{c|}{\textbf{LUT}}     & \textbf{Bitwidth} \\ \hline
TBCAS'23\cite{detection_ic}                  & \multirow{2}{*}{median}                                                     & \multicolumn{1}{c|}{\multirow{2}{*}{97.71}} & \multirow{2}{*}{98.68} & 1                               & \multicolumn{1}{c|}{\textgreater7500} & -                 \\ \cline{1-1} \cline{5-7} 
BioCAS'24\cite{l_sort}                       &                                                                             & \multicolumn{1}{c|}{}                       &                        & 1                               & \multicolumn{1}{c|}{1278}             & 525               \\ \hline
\multirow{2}{*}{TBCAS'23\cite{detection_ic}} & \multirow{2}{*}{median-of-median}                                           & \multicolumn{1}{c|}{\multirow{2}{*}{98.02}} & \multirow{2}{*}{97.34} & 6                               & \multicolumn{1}{c|}{468}              & 500               \\ \cline{5-7} 
                                             &                                                                             & \multicolumn{1}{c|}{}                       &                        & 1                               & \multicolumn{1}{c|}{2312}             & 500               \\ \hline
\textbf{This work}                           & \textbf{\begin{tabular}[c]{@{}c@{}}approx.\\ median-of-median\end{tabular}} & \multicolumn{1}{c|}{\textbf{97.52}}         & \textbf{94.73}         & \textbf{1}                      & \multicolumn{1}{c|}{\textbf{1600}}    & \textbf{104}      \\ \hline
\end{tabular}
\end{table}

\begin{figure*}[t]
\centering
\resizebox{.8\linewidth}{!}{
\begin{tikzpicture}[]
\foreach \signal [count=\y] in {
      {1,1,0,0},
      {127,63,0,0},
      {147,254,0,0},
      {2234,2235,2236,0},
    } {
    \ifnum\y=1
    \node at (0, \y+0.25) {valid};
    \node at (2.5, \y+4) {\color{red} \textcircled{1}};
    \node at (3.75, \y+4+0.03) {\color{green} \textcircled{2}};
    \node at (5, \y+4) {\color{blue} \textcircled{3}};
    \fi
    \ifnum\y=2
    \node at (0, \y+0.25) {amplitude};
    \fi
    \ifnum\y=3
    \node at (0, \y+0.25) {channel index};
    \fi
    \ifnum\y=4
    \node at (0, \y+0.25) {timestep};
    \fi
    \foreach \val [count=\c] in \signal {
        \draw (1.25*\c,\y) -- ++ (0.5, 0) -- ++ (0.25, 0.5) -- ++ (0.5, 0);
        \draw (1.25*\c,\y+0.5) -- ++ (0.5, 0) -- ++ (0.25, -0.5) -- ++ (0.5, 0);
        \ifnum\c=1
        \node at (1.25*\c+1.25, \y+0.25) {\color{red} \val};
        \fi
        \ifnum\c=2
        \node at (1.25*\c+1.25, \y+0.25) {\color{green} \val};
        \fi
        \ifnum\c=3
        \node at (1.25*\c+1.25, \y+0.25) {\color{blue} \val};
        \fi
    }
}
\node at (3,.5) {\emph{from peak detector (detected peaks)}};

\coordinate (start_value) at (6.25, 1.25);
\draw[->] (start_value) -- ++ (.75,0) -- ++ (0,-.5) -- ++ (.5,0) coordinate (value_0_end);
\draw[very thick] (value_0_end) ++(0, -0.25) rectangle ++(1, 0.5);
\node at ([xshift=0.5cm]value_0_end) {$=0$};
\coordinate (value_0_end) at ([xshift=1cm]value_0_end);
\draw[->] (value_0_end) -- ++ (.5,0) coordinate (value_0_end);
\draw[very thick] (value_0_end) ++(0, -0.25) rectangle ++(1.25, 0.5);
\node at ([xshift=0.625cm]value_0_end) {send?};

\draw[->] (start_value) ++ (.75,0) -- ++ (0,.5) -- ++ (.5,0) coordinate (value_1_end);
\draw[very thick] (value_1_end) ++(0, -0.25) rectangle ++(1, 0.5);
\node at ([xshift=0.5cm]value_1_end) {$=1$};
\coordinate (value_1_end) at ([xshift=1cm]value_1_end);
\draw[->] (value_1_end) -- ++ (.5,0) coordinate (value_1_end);
\draw (value_1_end) ++(0, -0.5) rectangle ++(1.25, 0.5);
\node at ([xshift=0.625cm,yshift=-0.25cm]value_1_end) {merge?};
\draw (value_1_end) ++(0, 0) rectangle ++(1.25, 0.5);
\node at ([xshift=0.625cm,yshift=0.25cm]value_1_end) {new?};
\draw[very thick] (value_1_end) ++(0, -0.5) rectangle ++(1.25, 1);

\coordinate (start_memory) at (12, 0.5);
\foreach \m in {0,...,7} {
    \draw[draw=black] (start_memory) ++(0,.5*\m) rectangle ++ (3,.5);
    \ifnum\m<6 
    \node at ([xshift=4cm, yshift=0.5*\m cm+0.25cm]start_memory) {Buffer \#\m};
    \fi
    \ifnum\m>6 
    \node at ([xshift=4cm, yshift=0.5*\m cm+0.25cm]start_memory) {Buffer \#15};
    \fi
}
\draw (start_memory) ++ (1,0) -- ++ (0,3);
\draw (start_memory) ++ (2,0) -- ++ (0,3);
\draw (start_memory) ++ (1,3.5) -- ++ (0,0.5);
\draw (start_memory) ++ (2,3.5) -- ++ (0,0.5);
\node at ([xshift=1.5cm, yshift=3.25cm]start_memory) {...};
\node at ([xshift=.5cm,yshift=4.33cm]start_memory) {ts.};
\node at ([xshift=1.5cm,yshift=4.33cm]start_memory) {ch.};
\node at ([xshift=2.5cm,yshift=4.25cm]start_memory) {amp.};
\node at ([xshift=.5cm,yshift=0.25cm]start_memory) {\textbf{2216}};
\node at ([xshift=.5cm,yshift=0.75cm]start_memory) {2221};
\node at ([xshift=.5cm,yshift=1.25cm]start_memory) {\textbf{2232}};
\node at ([xshift=.5cm,yshift=1.75cm]start_memory) {2233};
\node at ([xshift=.5cm,yshift=2.25cm]start_memory) {\textbf{2234}};
\node at ([xshift=.5cm,yshift=2.75cm]start_memory) {0};
\node at ([xshift=.5cm,yshift=3.75cm]start_memory) {0};
\node at ([xshift=1.5cm,yshift=0.25cm]start_memory) {\textbf{89}};
\node at ([xshift=1.5cm,yshift=0.75cm]start_memory) {317};
\node at ([xshift=1.5cm,yshift=1.25cm]start_memory) {\textbf{255}};
\node at ([xshift=1.5cm,yshift=1.75cm]start_memory) {64};
\node at ([xshift=1.5cm,yshift=2.25cm]start_memory) {\textbf{147}};
\node at ([xshift=1.5cm,yshift=2.75cm]start_memory) {0};
\node at ([xshift=1.5cm,yshift=3.75cm]start_memory) {0};
\node at ([xshift=2.5cm,yshift=0.25cm]start_memory) {\textbf{231}};
\node at ([xshift=2.5cm,yshift=0.75cm]start_memory) {142};
\node at ([xshift=2.5cm,yshift=1.25cm]start_memory) {\textbf{94}};
\node at ([xshift=2.5cm,yshift=1.75cm]start_memory) {124};
\node at ([xshift=2.5cm,yshift=2.25cm]start_memory) {\textbf{127}};
\node at ([xshift=2.5cm,yshift=2.75cm]start_memory) {0};
\node at ([xshift=2.5cm,yshift=3.75cm]start_memory) {0};
\draw[very thick] (start_memory) rectangle ++ (3,4);
\node at ([xshift=1.5cm,yshift=-.35cm]start_memory) {\textbf{Spike Bank}};

\coordinate (start_amp) at (6.25, 2.25);
\draw[->] (start_amp) -- ++ (2.75,0);
\coordinate (start_ch) at (6.25, 3.25);
\draw[->] (start_ch) -- ++ (2.85,0) -- ++ (0,-1);
\coordinate (start_ts) at (6.25, 4.25);
\draw[->] (start_ts) -- ++ (2.95,0) -- ++ (0,-2);

\draw [->] ([xshift=.5cm,yshift=4.5cm]start_memory) -- ++ (0,0.2) -- ++ (-2.5,0) -- ++ (0,-2.95);
\draw [->] ([xshift=1.5cm,yshift=4.5cm]start_memory) -- ++ (0,0.3) -- ++ (-3.6,0) -- ++ (0,-3.05);
\draw [->] ([xshift=2.5cm,yshift=4.5cm]start_memory) -- ++ (0,0.4) -- ++ (-4.7,0) -- ++ (0,-3.15);

\draw[->] (start_ts) ++ (.5,0) -- ++ (0,-4) -- ++ (2.6,0) -- ++ (0,0.25);
\draw[->] ([yshift=0.25cm]start_memory) -- ++ (-1.75,0);

\draw [->, red, ultra thick] ([xshift=1.25cm,yshift=0.25cm]value_1_end) -- ++ (0.75,0) -- node[midway, left] {\textcircled{1}} ++ (0,0.75) -- ++ (1,0);
\draw[draw=red,ultra thick] (start_memory) ++(0,2) rectangle ++ (3,.5);

\draw [->, green, ultra thick] ([xshift=1.25cm,yshift=-0.25cm]value_1_end) -- ++ (1,0) -- ++ (0,0.25) -- node[midway, above] {\textcircled{2}} ++ (0.75,0);
\draw[draw=green,ultra thick] (start_memory) ++(0,1) rectangle ++ (3,.5);

\draw [->, blue, ultra thick] ([yshift=-0.1cm]start_memory) -- node[midway, below] {\textcircled{3}} ++ (-2,0) -- ++ (0,-0.5);
\draw[draw=blue,ultra thick] (start_memory) ++(0,0) rectangle ++ (3,.5);

\draw[draw=black,ultra thick, rounded corners=5pt] (6.5,-0.25) rectangle ++ (10.5,6.5);
\node at (11.75,5.75) {\Large\textbf{Spike Locator}};

\draw[fill=white] ([xshift=0.625cm,yshift=-0.25cm]value_0_end) -- ++ (0,-.65) ++ (-0.1,0) rectangle ++ (0.2,-0.2) node[below] {\emph{to clustering module (located spikes)}};

\end{tikzpicture}
}
\caption{Hardware architecture of spike locator.}
\label{fig:locator}
\end{figure*}
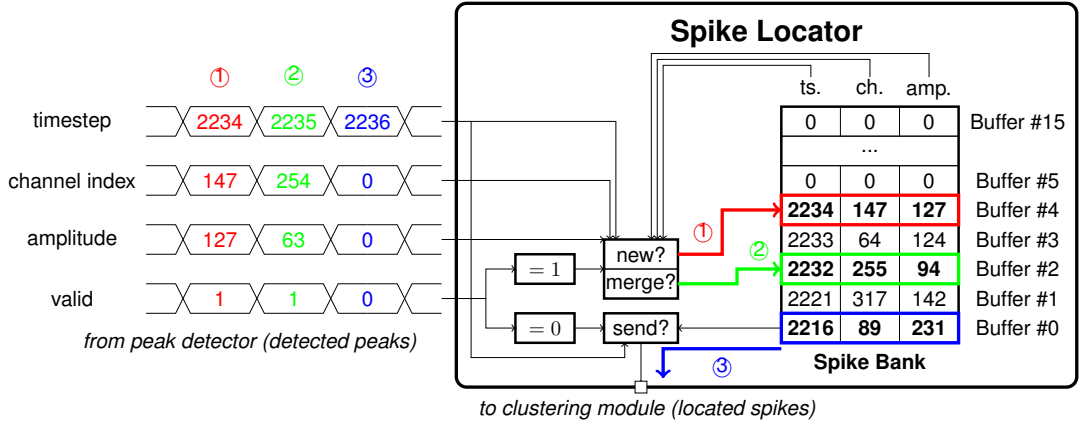

In our previous work~\cite{l_sort}, an incremental median calculator is proposed to further reduce the logical resources for finding medians, as shown in Fig.~\ref{fig:median_c}. Central to this proposal is to facilitate streamed samples. Since all the samples considered for the new median have already been sorted when calculating the median in the last time except the newest sample, it is possible to utilize the previously sorted sequence, along with removing the oldest point and inserting the new sample, to sort the new points. This method could simplify the median finding from comparing each point with each point to comparing only the new point with each point, i.e., from $O(n^2)$ to $O(n)$. 

The detailed hardware implementation of the proposed incremental median finding is shown in Fig.~\ref{fig:incrementaql_median}. Our proposed incremental method requires finding the oldest index and insert index, which are the indices for the oldest sample and the new sample, respectively. In our implementation, the previous points are stored in a RAM including their amplitude values and associated counters for tracking their temporal order. The counter of the new sample is set to $N-2$ and decreased by $1$ after each RAM access. When the value of the counter becomes $0$, it indicates that the corresponding sample is the oldest one of the stored points, thus acquiring the oldest index. On the other hand, the insert index is calculated by comparing the new sample with all previous points. These two indices are used for controlling multiplexers to sort the current points, which are written back to the RAM and indexed to find the median. This median is used to calculate the threshold, which is compared with the new sample to determine the presence of a peak.

The implementation of this incremental median calculator consumes only $1278$ LUTs~\cite{l_sort}. Also, because there is no modification on the median-finding algorithm, this implementation method can find the exact median instead of the approximate median, i.e., median-of-median. However, the memory utilization is not optimized. Moreover, because extra bits are required for storing the timestep information for each sample, additional memory is utilized. 

Considering that memory is often the primary consumer of hardware resources, a median calculator with a low memory footprint can reduce both the area and power consumption significantly. In this paper, we propose an incremental median-of-median calculator, as depicted in Fig.~\ref{fig:median_d}. The proposed architecture is composed of two separate stages of median buffers, each of which is similar to the design in Fig.~\ref{fig:median_c} with only four samples stored. The first buffer stage accepts the newest sample and provides the median of the five most recent samples. The second buffer stage accepts the median calculated by the first buffer stage and is updated every five timesteps, thus providing an approximate median-of-median calculation. As the acquired result is different from the median-of-median, we evaluated this scheme in real recordings to examine its performance in detection accuracy. Results show that this approximation has only \textless1\% and \textless3\% accuracy degradation in dataset 1 and dataset 2~\cite{dataset_neuropixel}, respectively. As for the hardware consumption, because only two sets of four comparators are needed for sorting the sequences, this design only consumes $1600$ LUTs. Moreover, because fewer points are involved in each median finding, fewer bits are required to mark the timesteps, and fewer samples need to be stored. Hence, the memory consumption is significantly reduced.

\subsection{Approximate Spike Localization}

The spike locator is responsible for grouping the detected peaks into spikes and localizing these spikes, as shown in Fig.~\ref{fig:locator}. To achieve these, a spike bank is implemented with $16$ buffers, containing up to $16$ ongoing spikes. For each spike, the peak with the highest amplitude along with its corresponding timestep and channel index are stored.

There are three different operations performed on this bank according to the input signal from the peak detector, for the sake of grouping peaks into spikes. When \emph{valid} from the detector is ‘1’, a peak is sent to the spike locator. This peak can be either the first detected peak of a new spike, or a succeeding peak of an ongoing spike. The comparisons between the peak and all existing spikes in the buffer are performed to examine the differences in timestep and channel index. If there is an ongoing spike both spatially and temporally proximate to the fed peak, the amplitudes between the matched spike and the new peak are further compared, and the timestep and channel index associated with the one with a higher amplitude is stored in the matched buffer. Otherwise, a new spike is generated in the buffer. For example, as demonstrated in Fig.~\ref{fig:locator}, \textcircled{1} has no matching on the existing buffer because there is no ongoing spike with a similar channel count, and therefore a new spike is generated at Buffer \#4. By contrast, \textcircled{2} is both spatially and temporally adjacent to the spike stored in Buffer \#2, but its amplitude is smaller than the stored amplitude and thus this buffer is not updated. When \emph{valid} from the detector is ‘0’, it indicates that there is no peak from the detector. The buffer will compare the timestep of the first buffer (Buffer \#0); if the current timestep is larger than the timestep stored in the first buffer by a threshold, the content (spike information) in the first buffer is sent out to the clustering module, as shown as \textcircled{3}.

As for the localization of the spikes, the conventional method utilizes center-of-mass (CoM) to calculate the geometric position of the spike source, which can be utilized as the spatial features for the succeeding clustering step. This method takes into account all adjacent channels around the central channel to the spike, which requires the storage of a time window of samples from all channels and consumes substantial memory consumption, thus degrading the hardware efficiency. In our previous design, we proposed the peak-based CoM, which utilizes only the detected peaks, therefore eliminating the access to amplitudes of all surrounding channels. Assuming that the probe is placed in XZ plane and the detected peaks are $P$, the peak-based CoM calculates the position of spike $[X_{spike}, Z_{spike}]$ as follows:
\begin{equation}
\begin{aligned}
    X_{spike} = \frac{\sum_{p\in P}amp_{p}x_{p}}{\sum_{p\in P}amp_{p}} \\
    Z_{spike} = \frac{\sum_{p\in P}amp_{p}z_{p}}{\sum_{p\in P}amp_{p}}
\end{aligned}
\end{equation}

However, this methodology requires the calculation and storage of several summations and products. These problems are more severe in scenarios when more buffers are required, i.e., when the system is up-scaled for high-channel-count probes. To make matters worse, the final results, i.e., the positions, are calculated as the division between two high-bitwidth summations. To overcome these obstacles, we propose a simplified solution in the localization process. Given that the pitches between recording sites are down to several micrometers, it is plausible to consider the position of the site with the highest amplitude, i.e., the central channel, as the approximate position for the spike source. By doing this, only the channel index is retained for localization, and the mathematical calculations are eliminated, thereby saving both computational and storage resources. We investigated the impact of this approximation on clustering accuracy, and results show that the difference is within 1\% in real high-density probe recordings.
\begingroup
\renewcommand{\arraystretch}{1.2}
\begin{table*}[t]
\centering
\caption{Performance of FPGA-based Multi-channel Spike Sorters}
\label{table:fpga_comp}
\setlength\tabcolsep{7pt}
\begin{tabular}{|clcccc|ccccc|}
\hline
\multicolumn{3}{|c|}{}                                                                                                                        & \multicolumn{3}{c|}{\textbf{Input Dataset}}                                                                            & \multicolumn{5}{c|}{\textbf{FPGA Utilization}}                                                                                                                                                        \\ \hline
\multicolumn{3}{|c|}{}                                                                                                                        & \multicolumn{1}{c|}{\textbf{Bitwidth}}          & \multicolumn{1}{c|}{\textbf{Sampling Rate}}     & \textbf{\#channel} & \multicolumn{1}{c|}{\textbf{LUT}}    & \multicolumn{1}{c|}{\textbf{FF}}     & \multicolumn{1}{c|}{\textbf{BRAM}} & \multicolumn{1}{c|}{\textbf{DSP}} & \textbf{Clock Freq.}                           \\ \hline
\multicolumn{3}{|c|}{TBCAS'2019~\cite{posort}}                                                                                                & \multicolumn{1}{c|}{-}                          & \multicolumn{1}{c|}{-}                          & -                  & \multicolumn{1}{c|}{$16472$}         & \multicolumn{1}{c|}{$8444$}          & \multicolumn{1}{c|}{$29$}          & \multicolumn{1}{c|}{$130$}        & $123$ MHz                                      \\ \hline
\multicolumn{3}{|c|}{Access'2020~\cite{zyon}}                                                                                                 & \multicolumn{1}{c|}{$12$-bit}                   & \multicolumn{1}{c|}{$18$ KHz}                   & $4096$             & \multicolumn{1}{c|}{$26444$}         & \multicolumn{1}{c|}{$28944$}         & \multicolumn{1}{c|}{$104$}         & \multicolumn{1}{c|}{$61$}         & $125$ MHz                                      \\ \hline
\multicolumn{3}{|c|}{TBME'2020~\cite{geo_osort_fpga}}                                                                                         & \multicolumn{1}{c|}{$16$-bit}                   & \multicolumn{1}{c|}{$20$ KHz}                   & $128$              & \multicolumn{1}{c|}{$17484$}         & \multicolumn{1}{c|}{$51674$}         & \multicolumn{1}{c|}{$98$}          & \multicolumn{1}{c|}{$60$}         & $200$ MHz                                      \\ \hline
\multicolumn{2}{|c|}{\multirow{5}{*}{BioCAS'2024~\cite{l_sort}}}                                        & \multicolumn{4}{c|}{Digital Filter}                                                                                                                          & \multicolumn{1}{c|}{$51$}            & \multicolumn{1}{c|}{$46$}            & \multicolumn{1}{c|}{$0.5$}         & \multicolumn{1}{c|}{$4$}          & \multirow{5}{*}{$3.6$ MHz}                     \\ \cline{3-10}
\multicolumn{2}{|c|}{}                                                                                  & \multicolumn{4}{c|}{Peak Detector}                                                                                                                           & \multicolumn{1}{c|}{$1237$}          & \multicolumn{1}{c|}{$84$}            & \multicolumn{1}{c|}{$5.5$}         & \multicolumn{1}{c|}{$0$}          &                                                \\ \cline{3-10}
\multicolumn{2}{|c|}{}                                                                                  & \multicolumn{4}{c|}{Spike Locator}                                                                                                                           & \multicolumn{1}{c|}{$1933$}          & \multicolumn{1}{c|}{$590$}           & \multicolumn{1}{c|}{$0$}           & \multicolumn{1}{c|}{$9$}          &                                                \\ \cline{3-10}
\multicolumn{2}{|c|}{}                                                                                  & \multicolumn{4}{c|}{Clustering Module}                                                                                                                       & \multicolumn{1}{c|}{$3387$}          & \multicolumn{1}{c|}{$2297$}          & \multicolumn{1}{c|}{$0$}           & \multicolumn{1}{c|}{$2$}          &                                                \\ \cline{3-10}
\multicolumn{2}{|c|}{}                                                                                  & \multicolumn{1}{c|}{total}          & \multicolumn{1}{c|}{$12$-bit}                   & \multicolumn{1}{c|}{$30$ KHz}                   & $120$              & \multicolumn{1}{c|}{$6513$}          & \multicolumn{1}{c|}{$3017$}          & \multicolumn{1}{c|}{$6$}           & \multicolumn{1}{c|}{$15$}         &                                                \\ \hline
\multicolumn{1}{|c|}{\multirow{12}{*}{\textbf{This work}}} & \multicolumn{1}{l|}{\multirow{6}{*}{c120}} & \multicolumn{4}{c|}{Digital Filter}                                                                                                                          & \multicolumn{1}{c|}{$73$}            & \multicolumn{1}{c|}{$84$}            & \multicolumn{1}{c|}{$0.5$}         & \multicolumn{1}{c|}{$4$}          & \multirow{6}{*}{\textbf{$\mathbf{3.6}$ MHz}}   \\ \cline{3-10}
\multicolumn{1}{|c|}{}                                     & \multicolumn{1}{l|}{}                      & \multicolumn{4}{c|}{Peak Detector}                                                                                                                           & \multicolumn{1}{c|}{$1600$}          & \multicolumn{1}{c|}{$194$}           & \multicolumn{1}{c|}{$2$}           & \multicolumn{1}{c|}{$1$}          &                                                \\ \cline{3-10}
\multicolumn{1}{|c|}{}                                     & \multicolumn{1}{l|}{}                      & \multicolumn{4}{c|}{Spike Locator}                                                                                                                           & \multicolumn{1}{c|}{$871$}           & \multicolumn{1}{c|}{$849$}           & \multicolumn{1}{c|}{$0$}           & \multicolumn{1}{c|}{$0$}          &                                                \\ \cline{3-10}
\multicolumn{1}{|c|}{}                                     & \multicolumn{1}{l|}{}                      & \multicolumn{4}{c|}{Clustering Module}                                                                                                                       & \multicolumn{1}{c|}{$96$}            & \multicolumn{1}{c|}{$134$}           & \multicolumn{1}{c|}{$0.5$}         & \multicolumn{1}{c|}{$0$}          &                                                \\ \cline{3-10}
\multicolumn{1}{|c|}{}                                     & \multicolumn{1}{l|}{}                      & \multicolumn{4}{c|}{UART}                                                                                                                                    & \multicolumn{1}{c|}{$126$}           & \multicolumn{1}{c|}{$70$}            & \multicolumn{1}{c|}{$0$}           & \multicolumn{1}{c|}{$0$}          &                                                \\ \cline{3-10}
\multicolumn{1}{|c|}{}                                     & \multicolumn{1}{l|}{}                      & \multicolumn{1}{c|}{\textbf{total}} & \multicolumn{1}{c|}{\textbf{$\mathbf{12}$-bit}} & \multicolumn{1}{c|}{\textbf{$\mathbf{30}$ KHz}} & $\mathbf{120}$     & \multicolumn{1}{c|}{$\mathbf{2564}$} & \multicolumn{1}{c|}{$\mathbf{1351}$} & \multicolumn{1}{c|}{$\mathbf{3}$}  & \multicolumn{1}{c|}{$\mathbf{5}$} &                                                \\ \cline{2-11} 
\multicolumn{1}{|c|}{}                                     & \multicolumn{1}{l|}{\multirow{6}{*}{c384}} & \multicolumn{4}{c|}{Digital Filter}                                                                                                                          & \multicolumn{1}{c|}{$74$}            & \multicolumn{1}{c|}{$90$}            & \multicolumn{1}{c|}{$0.5$}         & \multicolumn{1}{c|}{$4$}          & \multirow{6}{*}{\textbf{$\mathbf{11.54}$ MHz}} \\ \cline{3-10}
\multicolumn{1}{|c|}{}                                     & \multicolumn{1}{l|}{}                      & \multicolumn{4}{c|}{Peak Detector}                                                                                                                           & \multicolumn{1}{c|}{$1353$}          & \multicolumn{1}{c|}{$203$}           & \multicolumn{1}{c|}{$2$}           & \multicolumn{1}{c|}{$1$}          &                                                \\ \cline{3-10}
\multicolumn{1}{|c|}{}                                     & \multicolumn{1}{l|}{}                      & \multicolumn{4}{c|}{Spike Locator}                                                                                                                           & \multicolumn{1}{c|}{$1260$}          & \multicolumn{1}{c|}{$884$}           & \multicolumn{1}{c|}{$0$}           & \multicolumn{1}{c|}{$0$}          &                                                \\ \cline{3-10}
\multicolumn{1}{|c|}{}                                     & \multicolumn{1}{l|}{}                      & \multicolumn{4}{c|}{Clustering Module}                                                                                                                       & \multicolumn{1}{c|}{$113$}           & \multicolumn{1}{c|}{$148$}           & \multicolumn{1}{c|}{$0.5$}         & \multicolumn{1}{c|}{$0$}          &                                                \\ \cline{3-10}
\multicolumn{1}{|c|}{}                                     & \multicolumn{1}{l|}{}                      & \multicolumn{4}{c|}{UART}                                                                                                                                    & \multicolumn{1}{c|}{$158$}           & \multicolumn{1}{c|}{$70$}            & \multicolumn{1}{c|}{$0$}           & \multicolumn{1}{c|}{$0$}          &                                                \\ \cline{3-10}
\multicolumn{1}{|c|}{}                                     & \multicolumn{1}{l|}{}                      & \multicolumn{1}{c|}{\textbf{total}} & \multicolumn{1}{c|}{\textbf{$\mathbf{12}$-bit}} & \multicolumn{1}{c|}{\textbf{$\mathbf{30}$ KHz}} & $\mathbf{384}$     & \multicolumn{1}{c|}{$\mathbf{2658}$} & \multicolumn{1}{c|}{$\mathbf{1417}$} & \multicolumn{1}{c|}{$\mathbf{3}$}  & \multicolumn{1}{c|}{$\mathbf{5}$} &                                                \\ \hline
\end{tabular}
\end{table*}
\endgroup

\subsection{Clustering Module}
The clustering of the detected spikes are performed with O-Sort scheme~\cite{posort}, which mainly includes 1) merging the spike into an existing cluster or creating a new cluster and 2) updating the merged cluster and merging this cluster to other clusters if the updated cluster is similar to another cluster.

For conventional O-Sort coupling with FSDE features, the threshold for determining the merging of spike or cluster to cluster is calculated on-the-fly. By contrast, the localization-based features have a realistic meaning in biology, i.e., the positions of spike sources. The intervals among these sources can be estimated based on the probed neuronal area. Therefore, we use a programmable fixed threshold instead of calculating it dynamically to reduce resource consumption.

\begin{figure}[t]
  \centering
  \includegraphics[width=.8\linewidth]{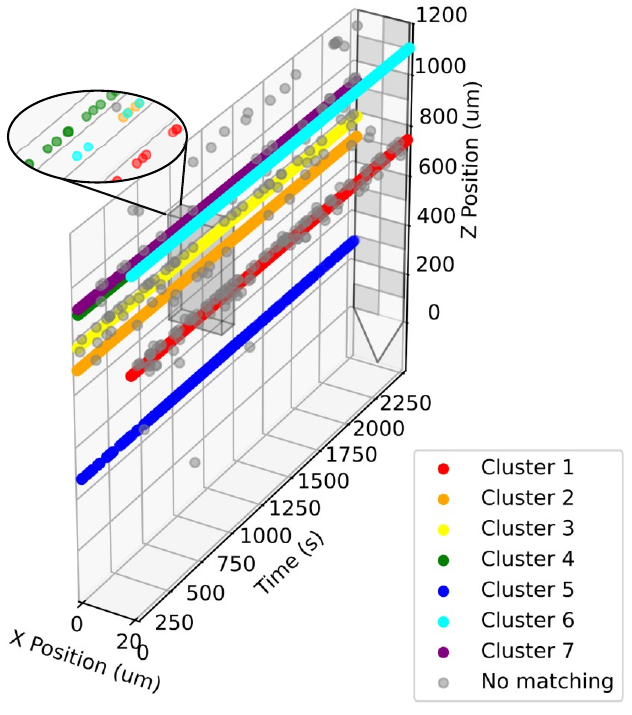}
  \caption{Spike clustering of Set 1 in \cite{dataset_neuropixel}. Clusters matched with the seven clusters provided in the ground truth are marked with non-gray colors, while others are marked with gray.}
  \label{fig:clustering}
\end{figure}
\section{Experimental Results}
\label{sec:experiments}

To assess the performance of the proposed L-Sort architecture, the prototype hardware is developed using SystemVerilog and implemented on both FPGA and ASIC. The datasets used in evaluating the sorting accuracy were recorded with Neuropixels probes, which were also used in examining~\cite{jssc_2023}. Because only partial recording sites ($120$ channels) are enabled during the recording of these datasets, the data are duplicated to match the full number of channels available on the Neuropixels probe, i.e., $384$ channels.

\subsection{FPGA Implementation}

The FPGA-based prototyping of the proposed L-Sort is conducted on the ZCU104 board, which is equipped with a Zynq UltraScale+ XCZU7EV MPSoC containing both FPGA and a quad-core ARM Cortex-A53 processor as PL and PS sides, respectively. The PS side is configured with the PetaLinux provided by Xilinx and communicates with the host PC through Ethernet for PL configuration and result visualization. The raw recording is saved on the SD card and read into DDR by PS before transmitting to the PL side for spike sorting. The synthesis, place, and route are performed using Xilinx Vivado 2024.1. The sorting results are shown in Fig.~\ref{fig:clustering}. Our design achieves $97.52\%$ and $97.23\%$ detection and classification accuracy, respectively.

The FPGA resource utilization of the design is shown in TABLE~\ref{table:fpga_comp}. We access the FPGA utilization, including look-up tables (LUTs), flip-flops (FFs), block RAMs (BRAMs), and digital signal processors (DSPs), and clock frequency required for achieving real-time processing for the proposed design supporting the processing of different numbers of channels. These results are compared with the existing state-of-the-art designs and our previously reported work~\cite{l_sort}.

We studied the hardware utilization of the proposed L-Sort with two different configurations with different numbers of channels, i.e., 120 channels (\emph{c120}) and 384 channels (\emph{c384}), respectively. It could be seen that both of our implementations consume significantly fewer hardware resources compared with other works. While others require massive calculations for calculating morphological features, which also associated with bulky storage for keeping the spike train during spike sorting, our localization-based method eliminates the requirement of keeping a whole time window for each spike throughout the feature extraction.

\begin{figure}[t]
  \centering
  \includegraphics[width=.85\linewidth]{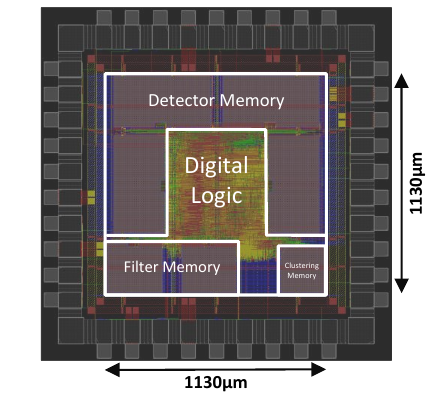}
  \caption{ASIC Layout of L-Sort.}
  \label{fig:dieshot}
\end{figure}

\begin{figure}[t]
    \centering
    \begin{subfigure}[b]{\linewidth}
        \centering
        \includegraphics[scale=0.45]{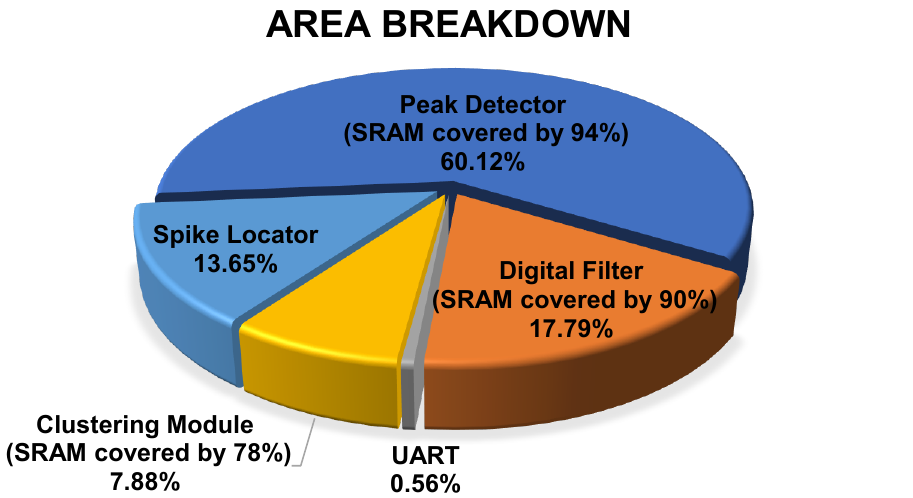}
        \caption{Area breakdown.}
        \label{fig:area_breakdown}
    \end{subfigure}

    \begin{subfigure}[b]{\linewidth}
    \vspace{10pt}
        \centering
        \includegraphics[scale=0.45]{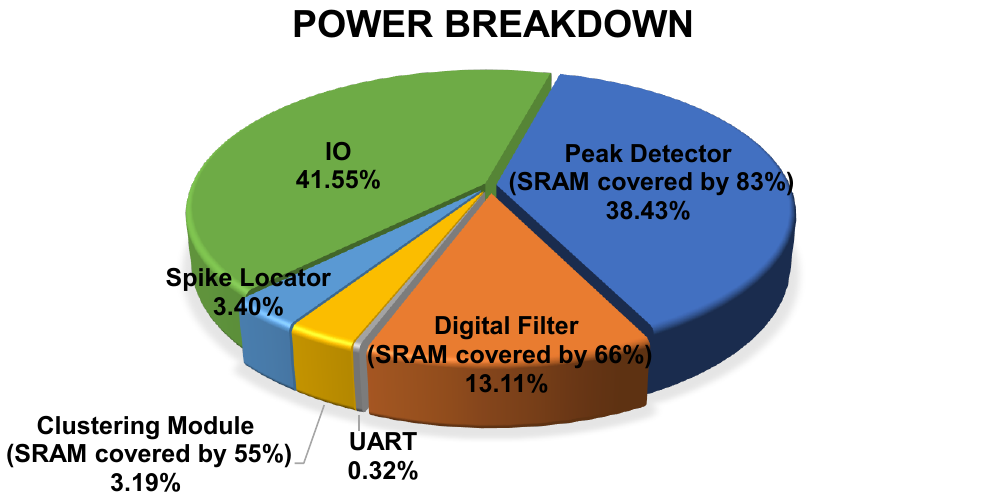}
        \caption{Power breakdown.}
        \label{fig:power_breakdown}
    \end{subfigure}

    \begin{subfigure}[b]{\linewidth}
    \vspace{10pt}
        \centering
        \scalebox{.9}{
            \renewcommand{\arraystretch}{1.3}
\setlength\tabcolsep{5pt}
\begin{tabular}{|c|c|c|c|}
\hline
Module    & \begin{tabular}[c]{@{}c@{}}Digital\\ Filter\end{tabular} & \begin{tabular}[c]{@{}c@{}}Peak\\ Detector\end{tabular} & \begin{tabular}[c]{@{}c@{}}Clustering\\ Module\end{tabular} \\ \hline
Width     & 24                                                       & 102                                                     & 9                                                           \\ \hline
Depth     & 384                                                      & 384                                                     & 384                                                         \\ \hline
Type      & 1r1w                                                     & 1r1w                                                    & single-port                                                 \\ \hline
Bandwidth & 48                                                       & 204                                                     & 9                                                           \\ \hline
Volume    & 9216                                                     & 39168                                                   & 3456                                                        \\ \hline
\end{tabular}
        }
        \caption{Memory breakdown in bits.}
        \label{fig:memory_breakdown}
    \end{subfigure}
    
    \caption{Details about resource consumption of the proposed L-Sort implemented on digital ASIC.}
    \label{fig:asic_breakdown}
\end{figure}

The first configuration (\emph{c120}) can process the same number of channels as our previously published design. It could be seen that the results achieved in this work demonstrate less hardware utilization in both logical and memory resources compared with the figures achieved in \cite{l_sort}. In our previous design, the feature extraction and spike clustering consumed the majority of the logical and computational resources, i.e., LUTs and DSPs. On the other hand, our current design utilizes a simpler method of calculating the locations and achieves a significant reduction in resource consumption. The numbers of LUTs utilized in both the spike locator and clustering module are decreased, because the summations used in CoM-based localizations are excluded and the determination for attributing the spikes to existing clusters is simplified.  Also, the DSPs used in the spike locator are eliminated as the dividers are no longer required.

To demonstrate the hardware utilization of the proposed design when scaling with the channel counts, the second configuration (\emph{c384}) is implemented to be capable of processing $384$ channels. The most obvious difference is the clock rate. To meet the real-time requirement, the clock frequency is set to the sampling rate multiplied by the clock cycle. Another major difference is the consumed storage space. However, in the Xilinx FPGAs from the Ultrascale+ series, the minimum depth of each BRAM is $512$ words. Considering that each word is allocated for keeping information of one channel, therefore the number of used BRAMs is only dependent on the bandwidth requirement, as long as the channel count does not exceed $512$ in FPGA-based implementations. As for the logical and computational resources, because they are shared among different channels through time-multiplexing, no duplications are required whereas the time closure can be met.

\subsection{ASIC Implementation}

\begingroup
\renewcommand{\arraystretch}{1.2}
\begin{table*}[t]
\centering
\caption{State-of-the-art On-chip Spike Sorting}
\label{table:sota}
\begin{threeparttable}
\setlength\tabcolsep{5pt}
\begin{tabular}{|cc|c|c|c|c|c|c|c|}
\hline
\multicolumn{2}{|c|}{Work}                                                                                      & \begin{tabular}[c]{@{}c@{}}ESSCIRC'2018\\ \cite{li2018esscirc}\end{tabular} & \begin{tabular}[c]{@{}c@{}}TVLSI'2019\\ \cite{tuan2019tvlsi}\end{tabular} & \begin{tabular}[c]{@{}c@{}}TBCAS'2021\\ \cite{han2021tbcas}\end{tabular} & \begin{tabular}[c]{@{}c@{}}TVLSI'2022\\ \cite{fereshteh2022tvlsi}\end{tabular} & \begin{tabular}[c]{@{}c@{}}TBCAS'2022\\ \cite{zeinolabedin2022tbcas}\end{tabular} & \begin{tabular}[c]{@{}c@{}}JSSC'2023\\ \cite{jssc_2023}\end{tabular} & \textbf{This work}                    \\ \hline
\multicolumn{1}{|c|}{\multirow{3}{*}{Algorithm}} & \begin{tabular}[c]{@{}c@{}}Spike\\ Detection\end{tabular}    & \begin{tabular}[c]{@{}c@{}}Absolute\\ thresholding\end{tabular}             & ICD                                                                       & \begin{tabular}[c]{@{}c@{}}Absolute\\ thresholding\end{tabular}          & \begin{tabular}[c]{@{}c@{}}Absolute\\ thresholding\end{tabular}                & NEO                                                                               & NEO                                                                  & \textbf{Median}                       \\ \cline{2-9} 
\multicolumn{1}{|c|}{}                           & \begin{tabular}[c]{@{}c@{}}Feature\\ Extraction\end{tabular} & \begin{tabular}[c]{@{}c@{}}Max-min\\ detection\end{tabular}                 & \begin{tabular}[c]{@{}c@{}}Integer\\ coefficient\end{tabular}             & FSDE                                                                     & Not applied                                                                    & \begin{tabular}[c]{@{}c@{}}Adaptive\\ filter\end{tabular}                         & Peak-FSDE                                                            & \textbf{Location}                     \\ \cline{2-9} 
\multicolumn{1}{|c|}{}                           & Clustering                                                   & \begin{tabular}[c]{@{}c@{}}1.5D Bayesian\\ boundary\end{tabular}            & \begin{tabular}[c]{@{}c@{}}Modified\\ K-means\end{tabular}                & \begin{tabular}[c]{@{}c@{}}Perturbed\\ K-means\end{tabular}              & \begin{tabular}[c]{@{}c@{}}CC-based\\ clustering\end{tabular}                  & Configurable                                                                      & Geo-OSort                                                            & \textbf{OSort}                        \\ \hline
\multicolumn{2}{|c|}{Number of channels}                                                                        & 96                                                                          & 128                                                                       & 4                                                                        & 64                                                                             & 16                                                                                & 384                                                                  & \textbf{384}                          \\ \hline
\multicolumn{2}{|c|}{Dataset}                                                                                   & -                                                                           & Quiroga                                                                   & Quiroga                                                                  & Quiroga                                                                        & Quiroga                                                                           & Neuropixel                                                           & \textbf{Neuropixel}                   \\ \hline
\multicolumn{2}{|c|}{Accuracy}                                                                                  & -                                                                           & 72\%                                                                      & 93.2\%                                                                   & 85\%                                                                           & 94.1\%                                                                            & 97.7\%                                                               & \textbf{97.2\%}                       \\ \hline
\multicolumn{2}{|c|}{Technology}                                                                                & 180nm                                                                       & 65nm                                                                      & 180nm                                                                    & 180nm                                                                          & 22nm                                                                              & 22nm                                                                 & \textbf{180nm/22nm}                   \\ \hline
\multicolumn{2}{|c|}{Core Voltage}                                                                              & 0.32V                                                                       & 0.54V                                                                     & 1.5V                                                                     & 1.8V                                                                           & 0.5V-0.8V                                                                         & 0.59V                                                                & \textbf{1.8V}                         \\ \hline
\multicolumn{2}{|c|}{\begin{tabular}[c]{@{}c@{}}Number of Bits of\\ Input Data\end{tabular}}                    & 8-bit                                                                       & 9-bit                                                                     & Analog input                                                             & 8-bit                                                                          & 9-bit                                                                             & 12-bit                                                               & \textbf{12-bit}                       \\ \hline
\multicolumn{2}{|c|}{\begin{tabular}[c]{@{}c@{}}Sampling Rate of\\ Input Data\end{tabular}}                     & 30kHz                                                                       & 25kHz                                                                     & 24kHz                                                                    & 25kHz                                                                          & 25kHz                                                                             & 30kHz                                                                & \textbf{30kHz}                        \\ \hline
\multicolumn{2}{|c|}{Power per channel}                                                                         & 0.006$\mu$W                                                                 & 0.175$\mu$W                                                               & 4.68$\mu$W                                                               & 1.74$\mu$W                                                                     & 2.79$\mu$W                                                                        & 1.78$\mu$W                                                           & \textbf{71.03$\mu$W/0.314$\mu$W\dag}  \\ \hline
\multicolumn{2}{|c|}{Area per channel}                                                                          & 0.019mm$^2$                                                                 & 0.003mm$^2$                                                               & 1.023mm$^2$                                                              & 0.047mm$^2$                                                                    & 0.014mm$^2$                                                                       & 0.0013mm$^2$                                                         & \textbf{0.0033mm$^2$/4.9254e-5mm$^2$\dag} \\ \hline
\end{tabular}
\begin{tablenotes}
    \item[\dag] normalized to 22 nm with 0.7V voltage supply using scaling factors from~\cite{scale}.
\end{tablenotes}
\end{threeparttable}
\end{table*}
\endgroup

We also implemented our design in ASIC using a standard 180 nm CMOS technology, followed by post-place-and-route (post-PnR) simulations to assess the switching activity and power consumption. The synthesis and PnR are performed by Cadence\textsuperscript\textregistered\ Genus and Cadence\textsuperscript\textregistered\ Innovus, while the post-PnR simulations are conducted with Synopsys\textsuperscript\textregistered\ VCS, whose results are used to generate the switching activity file (.saif) for acquiring a more precise power consumption in Cadence\textsuperscript\textregistered\ Voltus. The layout of our implemented design is shown in Fig.~\ref{fig:dieshot}, where the on-chip SRAMs used in this design are generated by ARM\textsuperscript\textregistered Artisan Memory Compiler. The core area is $1.13\times1.13$ mm$^2$ ($1.2769$ mm$^2$), and the total power consumption including IOs is $27.28$ mW.

The breakdown of the implemented chip in terms of area and power consumption is shown in Fig.~\ref{fig:area_breakdown} and Fig.~\ref{fig:power_breakdown}, respectively. The IOs consume substantial power, which is limited to the technology utilized in this implementation, as the IOs are powered up to $5$ volts in the standard cells. Indicatively, the power utilization can be easily optimized further by embracing modern fabrication technology. Other than IOs, the peak detector is the dominant module with respect to both area and power. This is because the spike detection is an always-on operation regardless of the presence of spikes and the relatively high bandwidth of the utilized SRAM compared with other modules, as shown in Fig.~\ref{fig:memory_breakdown}. However, compared with the $76.6$ kB SRAM utilized for spike detection and feature extraction in~\cite{jssc_2023}, this design only consumes less than $5$ kB SRAM for spike detection, as well as a spike bank with $104$ Byte memory in spike locator, which is implemented as registers considering the requirement of accessing all buffers simultaneously.

We also compare the ASIC implementation results with other state-of-the-art spike sorters implemented on ASICs, as shown in TABLE~\ref{table:sota}. To demonstrate the performance of the proposed L-Sort under more advanced technology nodes, the area and power consumption are normalized to 22 nm and 0.7 V voltage supply using the scaling factors provided in~\cite{scale}, which have been widely used to compare the designs using different technology nodes~\cite{kassiri2017rail,jang20191}. The work~\cite{jssc_2023} utilized the geometric information to reduce the number of compared clusters during classification and was evaluated using the same Neuropixels dataset with this work. However, the features used in their design still utilized the derivative features calculated with the whole waveform, leading to an input buffer of $76.6$ kB consuming 39.3\% area in the whole chip. In contrast, our design optimized the memory consumption by using spatial features, reducing the memory volume for storing spike information. Hence, a significant improvement in area utilization has been achieved. As for the power consumption, the always-on detection is achieved simply with few comparators and multiplexers, thanks to the proposed incremental median-of-median calculator. Moreover, the approximate spike localization excluding multiplications and divisions further optimizes the power and area utilization. After normalization to the same technology node, i.e., 22 nm, our design achieves $96.2$\% area reduction and $82.4$\% comparing with ~\cite{jssc_2023}. Note that the SRAMs occupy $79.0$\% area and $42.5$\% power consumption, along with the IOs consume $41.5$\% power, which together contribute to the majority of the hardware consumption. As both SRAMs and IOs are general and standard elements in digital ASICs, they can be easily optimized with advanced technology nodes and benefit from advancements in corresponding automated generation tools.

\section{Conclusion}
\label{sec:conclusion}

In this paper, we propose L-Sort, which aims at improving the hardware efficiency of on-chip multi-channel spike sorting utilizing efficient median-based detection and localization-based clustering. On the one hand, the proposed incremental median-of-median calculation scheme reduces both the volume and access of memory significantly with minimal compromise in sorting accuracy. On the other hand, spike localization eliminates the requirement for keeping the whole spike waveforms during feature extraction. Moreover, the proposed approximate spike localization further excludes the multiplications and divisions in the original center-of-mass method and simplifies the clustering process, which is also evaluated to demonstrate its impact on sorting accuracy. By incorporating all these techniques, we first tested our design on FPGA to assess its performance. Compared with the other FPGA-based designs, this work achieved a significant reduction in resource utilization. Besides, we also implemented the design on ASIC with 180 nm technology and compared the post-PnR results with existing ASIC designs. Results show that our design demonstrates roughly $\times 10$ better area and power efficiency with similar accuracy after normalized to 22 nm technology, compared with the state-of-the-art designs evaluated with the same dataset.

\bibliographystyle{IEEEtran}
\bibliography{l-sort}

\end{document}